\documentclass[nohyper,12pt,letterpaper]{JHEP}
\usepackage{graphicx}

\newfont{\frak}{eufm10 scaled 1200}

\newfont{\Bbb}{msbm10 scaled 1200}     
\newcommand{\mathbb}[1]{\mbox{\Bbb #1}}
\DeclareSymbolFont{AMSa}{U}{msa}{m}{n}
\DeclareSymbolFont{AMSb}{U}{msb}{m}{n}
\let\Box\relax
\DeclareMathSymbol{\Box}{\mathord}{AMSa}{"03}

\def \eqn#1#2{\begin{equation}#2\label{#1}\end{equation}}

\title{Holographic Cosmology 3.0\footnote{This is the $\beta$ test version of holographic cosmology.  While it is
unlikely to crash your hardware when you think about it, no warranties, either explicit or implied, are provided
for this product.}}

\author{T. Banks \\
   Department of Physics and Astronomy - NHETC\\
   Piscataway, NJ 08540\\
   and\\
   Department of Physics, SCIPP\\
   University of California, Santa Cruz, CA 95064\\
E-mail: \email{banks@scipp.ucsc.edu}}

\author{ W. Fischler,  \\
     Department of Physics\\
     University of Texas, Austin, TX 78712\\
E-mail: \email{fischler@physics.utexas.edu}}

\abstract{We present a new version of holographic cosmology, which is compatible with present observations.  A
primordial $p=\rho$ phase of the universe is followed by a brief matter dominated era and a brief period of
inflation, whose termination heats the universe.  The flatness and horizon problems are solved by the $p=\rho$
dynamics. The model is characterized by two parameters, which should be calculable in a more fundamental approach
to the theory.  For a large range in the phenomenologically allowed parameter space, the observed fluctuations in
the cosmic microwave background were generated during the $p=\rho$ era, and are exactly scale invariant.  The
scale invariant spectrum cuts off sharply at both upper and lower ends, and this may have observational
consequences. We argue that the amplitude of fluctuations is small but cannot yet calculate it precisely.    }

\keywords{Holography, Inflation, Cosmology}

\preprint{\hepth{} \\RUNHETC-2003-32 \\SCIPP-03/10\\ UTTG-05-03}
\begin{document}


\section{Introduction - A Most Entropic Beginning ...}
A little over two years ago, we introduced a new approach to cosmological initial conditions called holographic 
cosmology\cite{holocosmo}.   The basic principle on which it was based is the holographic entropy bound
\cite{fsb}\cite{g} .   In a Big Bang cosmology, the bound implies a finite entropy for any causal cone (the causal
past
 of a finite point in space-time), which decreases to zero as we approach the initial singularity.
We interpreted this entropy as the entropy of the maximally uncertain density matrix for measurements
done inside the causal cone, a conjecture with several attractive features.

First of all, it introduces an element of locality into a fundamentally holographic theory:  a finite set of 
degrees of freedom is associated with a local region of space-time.   Secondly, the conjecture provides a
natural, nonsingular, interpretation for the initial singularity:  the (backward) evolution of the universe stops
when the particle horizon can support only a single qubit of information.  It is clear that although smooth
geometrical concepts are not appropriate in this regime, the quantum mechanical description of it is completely
finite.  Following through on this idea leads to a general framework for quantum cosmology, which we review in
Appendix A.  A third interesting feature of this conjecture is that the formalism automatically introduces an
arrow of time.  Time evolution (as seen by a given observer) is constrained in such a way that at early times,
only degrees of freedom which are within the particle horizon, are correlated by the dynamics.  This introduces a
direction of time:  as we go into the future, more degrees of freedom are correlated with each other.

This conjecture enabled us to address the question of the Initial Quantum State\cite{bfmcosmo}.  We argued 
that this question was inextricably tied up with the Problem of Time.    We will present an exegesis of
our view of the problem of time in Appendix A.    Here we only note that there are two possible hypotheses about
the nature of quantum gravity near an initial singularity\footnote{A third hypotheses is that every Big Bang
should be viewed as emerging from a quantum mechanically smoothed Big Crunch, with the fundamental observables
defined as ``S-matrix" elements relating states in the  asymptotic future of the Bang to those in the asymptotic
past of the Crunch.   There is also a ``cyclic" version of this idea.}.   One is that there is some special time
evolution operator, defining the quantum analog of a preferred coordinate system, which describes physics near
the Big Bang.  If this is the  case, one must also ask about the choice of initial state.   We prefer the
hypothesis that there is no special evolution operator.   Rather the early evolution should be a random  sequence
of random operators, drawn from some distribution, and subject only to the general consistency conditions of
quantum cosmology, described in the appendix.   In this case, the initial state can  be subsumed into the choice
of time evolution operator.  This approach to initial conditions is inherently statistical.   That is, if we want
it to be predictive, we must rely on universal properties of large random evolution operators.

The mathematics of this new form of quantum cosmology is difficult and new, and our progress in understanding 
it has been slow.   However, we have developed a heuristic picture of random quantum initial conditions, which
allowed us to elaborate a full semi-classical cosmology, in which the early universe is dominated by a $p=\rho$
fluid.   In the present paper we will make several modifications of this cosmological picture.   The most
important  modifications have to do with the exactly scale invariant fluctuation spectrum predicted by our
model.   In
\cite{holocosmo} we derived this fluctuation spectrum using quantum field theory in the $p=\rho$  background. 
This is inconsistent with our argument (see below) that this background has no inhomogeneous excitations.   In the
present paper we will rederive the spectrum using only a conformal symmetry of this background and some falloff
conditions on probability distributions. We find a scale invariant fluctuation spectrum over a limited range of
comoving scales, ranging from $k = 1$ to $k = M^{-2/3}$, where $M$ is the physical size of the horizon at the end
of the
$p=\rho$ era\footnote{Here and henceforth we work in Planck units.}.  We have found that even if we introduce
correlations between fluctuations on much larger scales (for example, through initial conditions), that scale
invariance does not extend beyond this limited range.  On larger scales the fluctuations are highly suppressed
relative to the scale invariant distribution (the spectrum is blue).  In our previous paper we found that $M$ was
bounded by phenomenological considerations and that this range of scales was consequently too small to account
for the observed CMB fluctuations.  We briefly considered, and too hastily discarded, the possibility of
stretching these scales by a period of relatively late inflation.  This option is quite natural in the context of
string theory, where moduli provide attractive candidates for the inflaton field.  Naive dimensional analysis in
a wide class of string compactifications suggests a small number of e-folds is natural.  At the level of that
analysis, $N_e \sim 3\pi^2$ could easily be considered small.  We will find that our scenario requires between
$17$ and $32$ e-foldings, depending on the value of the fundamental parameter $M$.  In principle, this parameter
is calculable, but phenomenology already puts stringent constraints on it.   The new version of holographic
cosmology (version 3.0) is not as constrained as the one in our original paper (which was not only falsifiable,
but actually false). For a large range of our parameter space, we are still forced to invoke low energy
baryogenesis ({\it e.g.} the Affleck-Dine mechanism \cite{ad}) and axion dark matter.  The only prediction of our
old model which definitely does not survive is the relic density of very heavy magnetically charged black holes.

 The first element of 
our heuristic cosmological picture was the observation of \cite{fs} that a perfect fluid
Friedmann-Robertson-Walker (FRW)
 cosmology, with flat spatial sections and equation of state $p=\rho$ can saturate the holographic entropy bound
for all times.   We argued \cite{bfmcosmo} that this  implied that the coarse grained picture of the early
universe, with random initial conditions, was automatically homogeneous isotropic and spatially flat. The entropy
bound implies that there can be no inhomogeneous fluctuations of a perfect $p=\rho$ fluid saturating the bound. 
{\it Note that this implies that quantum field theory is not a good description of fluctuations around this
background}.  This is a key fact, and one which we have found extraordinarily hard to explain to colleagues
steeped in the lore of local field theory.  We repeat it for emphasis: a HOMOGENEOUS $p=\rho$ fluid saturates the
entropy bound.  This means that fluctuations in this fluid must be changes in its quantum state which do not
change its local energy and pressure (just as fluctuations in the state of a black hole do not change its
macroscopic geometry).  A clue to how the quantum mechanics of this peculiar fluid might work may be found in the
Appendix.

 Curved versions of the $p=\rho$ cosmology are either unstable to local collapse (positive
curvature), or do not saturate the entropy bound (negative curvature). We argued that the positively curved
regions would just be viewed as part of the black hole fluid in a larger, flat $p=\rho$ background.

To proceed, we needed a mechanical model for the $p=\rho$ fluid.  We argued that it was a dense black hole fluid.   The argument is simple and bears repetition:   in $d$ space-time dimensions, a fluid of black holes with number density $n$
has entropy and energy densities given by
$$\sigma = A_{d-2} R_S^{d-2} n$$
$$\rho = R_S^{(d-3)} n,$$
where $A_{d-2}$ is the area of a unit $d-2$ sphere, and the equations, like all equations in this paper, are
 written in Planck units.
If the black holes have an average separation $k R_S$ then
$$\sigma = A_{d-2} k^{-{{d-1}\over 2}} \sqrt{\rho}$$.
For an appropriate choice of $k$, this saturates the entropy bound.  Thermodynamics then implies that if this
system is in  equilibrium, its equation of state is $p=\rho$.  Note that a homogeneous minimally coupled scalar
field (more generally, a $\sigma $ model) also has an energy momentum tensor satisfying this relation,
but does not have an extensive entropy, and so is not a good model for the primordial cosmic fluid.

Heuristically, the fluid of black holes will indeed be in equilibrium.   A given black hole saturates the
entropy bound in its particle horizon volume.  As the universe expands, black holes in disjoint horizon 
volumes merge to form larger black holes, which fill up the new horizon.   The pressure results from the rapid
expansion of the black hole horizon as black holes merge, which follows from the negative specific heat of black
holes.   Thus, we believe the dense black hole fluid is a stable steady state solution of the equations of
quantum cosmology.   

It should be emphasized that we are using the words {\it black hole} in a heuristic manner, which should
not be taken to imply particular forms of space-time metric, or related to ``the complement of the causal
past of null infinity".  Rather, we believe that thirty years of work on quantum gravity have established 
the existence of localized excitations of the theory with the energy, size, and degeneracy relations of black
holes in asymptotically flat space-time.  The rigorous justification of our heuristic picture should be sought,
not in relations to the definition of black holes in classical GR, but in the quantum cosmological framework
outlined above. We will outline our attempt to create a mathematical quantum mechanical model of the dense black
hole fluid in the Appendix.

We will conjecture one more property of the fluid in order to complete our analysis.   Every FRW cosmology with
 a single component equation of state, possesses a conformal Killing vector.   
Typically, the physics of the fluid driving the expansion of the universe is not invariant under this
conformal transformation.   For example, if $p={1\over 3} \rho$  the radiation fluid was hotter at early times.  
It is clear that corrections to its effective Lagrangian will become important in the hot early universe, and
conformal invariance will be destroyed.   By contrast, the ever merging fluid of black holes, does appear to be a
self similar system, and we will conjecture that this is the case.   This will be the crucial tool in our
argument for scale invariant density perturbations.   Again, the justification for this conjecture should be
sought in a true quantum mechanical model of the dense black hole fluid (perhaps the one we describe in the
Appendix).

If the above conjectures are confirmed, we will have shown that a generic initial condition
 for the universe evolves into a stable, dense black hole fluid.   This is not a very
interesting system.  At any given time, all of its degrees of freedom are in intense
interaction, which brings them into equilibrium.   Local physics is non-existent, and there are no complex
structures.  This motivates our consideration of atypical initial conditions, which are less entropic than
the black hole fluid, but have the maximum entropy consistent with the development of a ``normal" region
of the universe.  In the next section we devote some time to explaining what we understand about the nature
of such initial conditions.   We then present a new derivation of the scale invariant fluctuation spectrum.
Section 3 is devoted to a discussion of the late inflationary stage, and to the bounds on the various parameters
which enter our analysis.  In the conclusions, we contrast our model with conventional inflationary models.

This paper is based on lectures given by T. Banks at the Nobel Symposium, Sigtuna Stiftelsen, Sweden, June 14 -
19, 2003  and by W. Fischler at the Conference on String Theory and Cosmology, KITP, UCSB, Santa Barbara, CA ,
October 20-24, 2003.  The paper was written while both authors were resident at KITP during the String theory and
Cosmology Program.  We thank the staff and faculty of KITP for providing a stimulating environment for our work.

\section{Normal, Universe}

In our previous paper we imagined the evolution of a normal universe began from a sprinkling of droplets
within the $p=\rho$ fluid, in which black holes less than the horizon size were found.  This is a less
entropic initial condition than the homogeneous $p=\rho$ fluid.  We argued that if, in a region of linear size
10-100 times the horizon size, we had a collection of such non-maximal black holes, then they would
evolve as a $p=0$ gas, and then evaporate into radiation.  We argued that these regions would grow like a
universe which is eventually radiation dominated.

We have realized that there was an error in this analysis.  A sphere of normal fluid embedded in a $p=\rho$
background must satisfy the Israel junction conditions \cite{israel}.  The transverse component of this condition
involves the stress tensor on the interface, about which we know little, but the continuity of the geometry
must be satisfied.   In order to apply this condition, we must synchronize the times in the two FRW cosmologies.
We do this by using the equal area slicing which we introduced in our previous paper.  Up to a factor of order
one, this requires the cosmic times to be equal.  Consider a spherical interface with fixed coordinate size, $L$,
in the $p=\rho$ comoving coordinate system. The interior of this sphere is filled with normal fluid. Let $R(t)$ be
the coordinate size of this interface in the normal region, which we take to be radiation dominated.
The geometry will be continuous if (we work in four dimensions)
\eqn{isr}{t R^2 (t) = t^{2/3} L^2}
which implies (as we have anticipated in our notation) that the coordinate size of the sphere in the radiation
dominated region must shrink like $t^{- {1\over 6}}$.  Thus the junction condition implies, not only that the
particle horizon in the normal region cannot grow, but that the region actually shrinks.
Something similar would happen for any equation of state less stiff than $p=\rho$.
However, if the geometry in the normal region asymptotes to de Sitter space, then another embedding is required.
We cannot maintain the equal area slicing in FRW coordinates, since the dS horizon asymptotes to a finite
physical size.  Instead we embed the static region of dS space into the interior of a coordinate ball with
radius $L(t)
\sim t^{-{1\over 3}}$ in the $p=\rho$ fluid. 

In order to avoid the problem of recollapse of the normal region, we must choose a more complicated initial
condition.  One possibility would be to choose an initial normal region whose coordinate size was much larger
than our current horizon.  This is obviously a very low entropy situation.   However, we believe that there are
much more probable initial configurations which are able to grow within the $p=\rho$ fluid.  Consider a region
constructed by gluing together spheres of radius $L$ (there is no reason to take them all of the same radius).  
This sort of Tinker Toy (which we will henceforth call a fractal - without implying any scale invariant
properties) can fill up a finite fraction $\epsilon$ of the volume of a much larger sphere.  The latter sphere
has to have a coordinate size at least of order our current horizon volume\footnote{This would be sufficient if
our current universe asymptotes to dS space.  If our horizon is required to expand eternally, then the fractal
must extend to infinity.}.  A sufficiently complicated fractal can avoid the collapse to a
$p=\rho$ state, since collapse certainly does not occur for $\epsilon = 1$.  For $\epsilon = 1$ the entire volume
that we will ever observe is taken to be normal from the beginning.  We do not have the tools to determine the
optimal shape of the fractal.  Clearly, the most probable initial conditions, which give a normal universe,
will have  the smallest value of $\epsilon$ for which recollapse is avoided.  An artist's conception of the
initial state is shown in figure 1.

\includegraphics[width=5in]{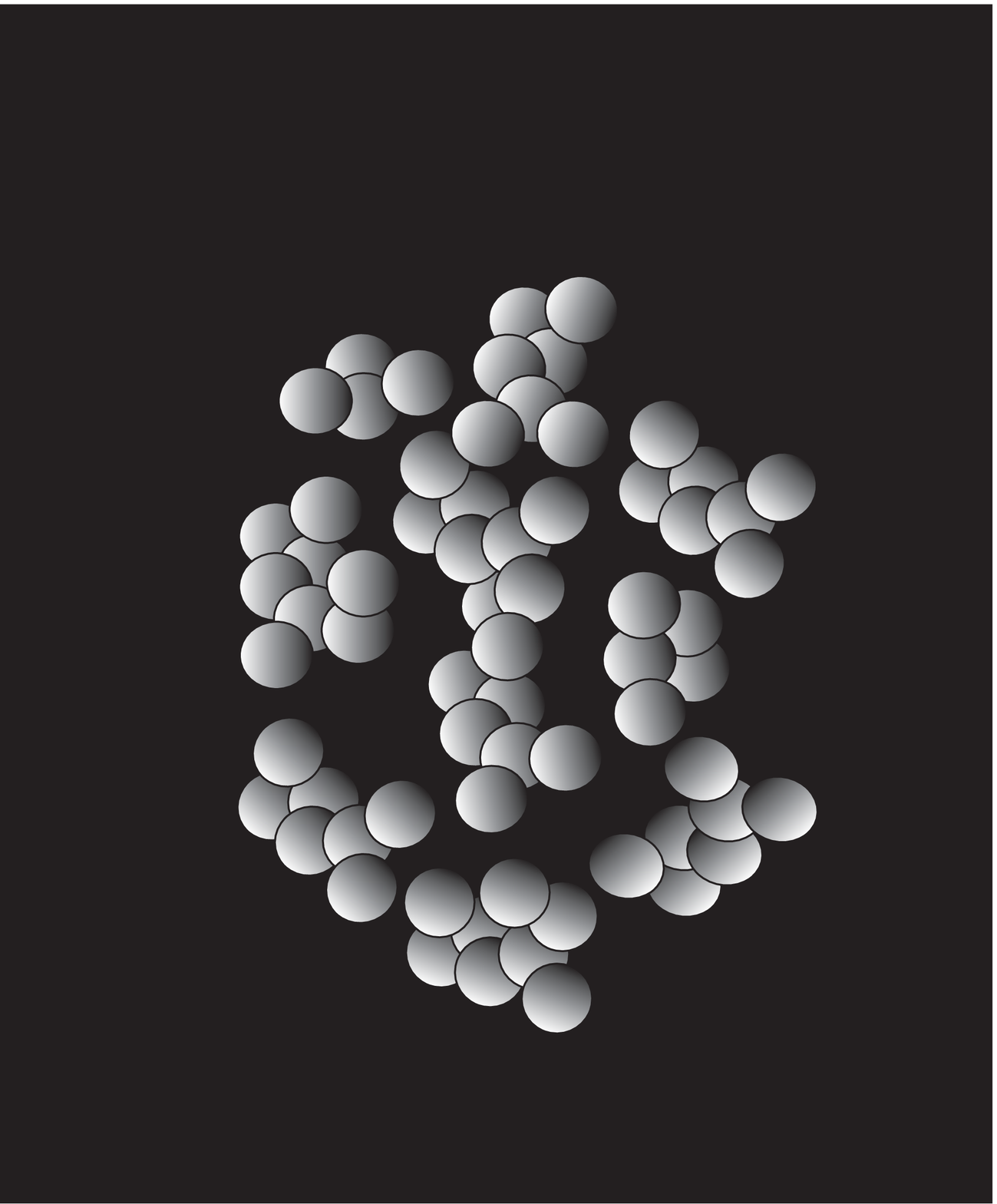}

\centerline{Figure 1: The Initial ``Fractal" Distribution of Normal Regions}
\centerline{\ \ \ \ \ In The Dense Black Hole Fluid}

The parameter $\epsilon$ is the crucial determinant of the subsequent history of the universe.  As time goes on,
the ratio of the volume in normal regions, to that in $p=\rho$ regions, grows like $t^{1/2}$.   Thus for $t > T =
{1\over \epsilon^2}$, the volume is dominated by normal regions. The artist's view of the universe now resembles
Fig. 2.  As seen from the normal regions, the $p=\rho$ regions trapped in the interstices of the fractal, behave
like black holes.  The average mass of these black holes is $M = T$.

\includegraphics[width=5in]{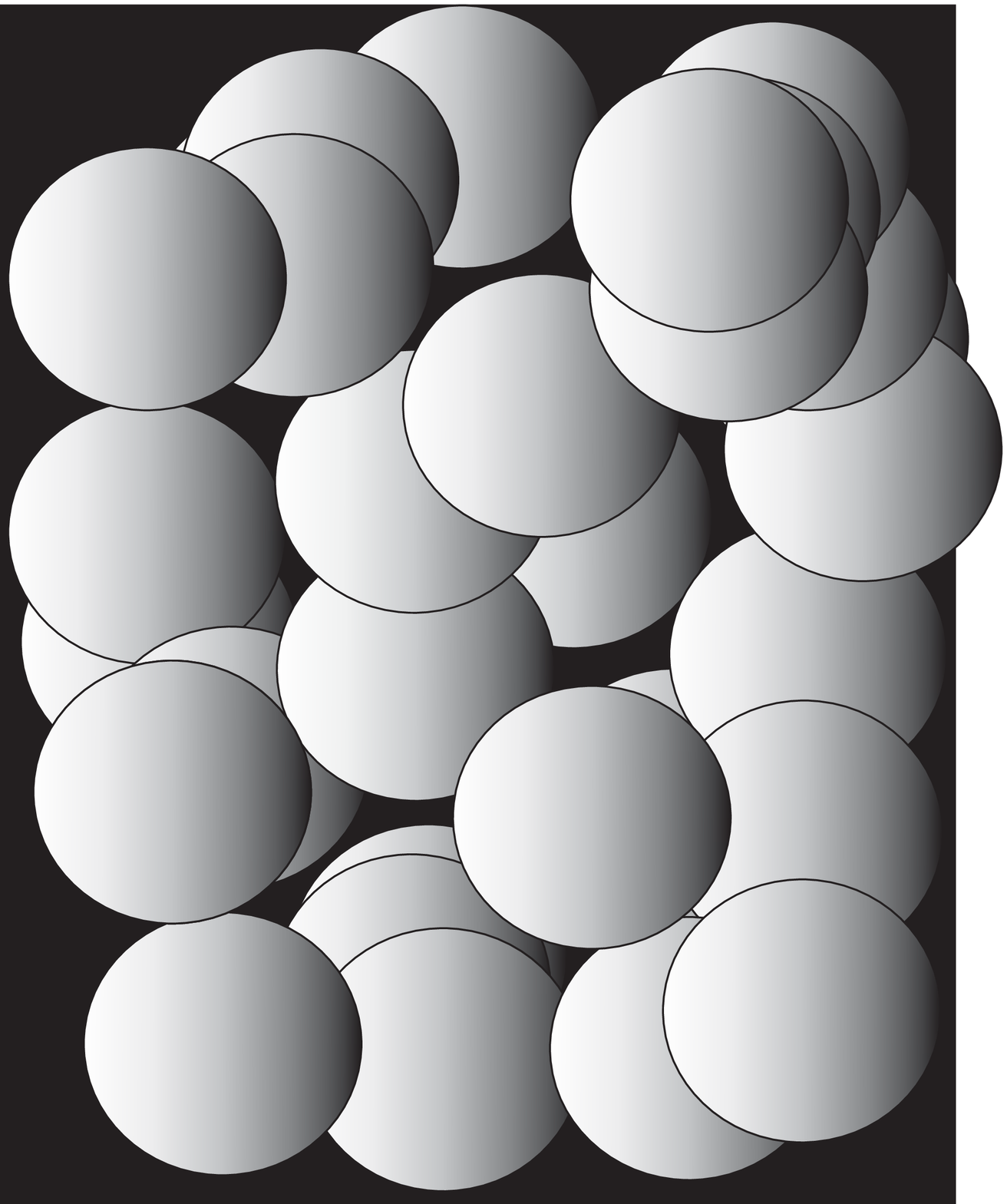}

\centerline{Figure 2: The Normal Regions Begin To Dominate The Volume of The Universe}

After this transition, the black holes cease to grow (and in fact begin to decay by Hawking radiation).  They
quickly come to dominate the energy density of the universe.  During this period the scale factor grows like
$t^{2/3}$.  Note that, at this point, we have explained the homogeneity, isotropy and flatness of the FRW
universe, without recourse to inflation.  The original $p=\rho$ background possesses these properties for
entropic reasons \cite{holocosmo}.  The interior of the fractal is locally flat for the same
reason.  The structure of the fractal, and the fluctuations in the sizes of the submaximal black holes inside it,
are the origin of density fluctuations.  As we will see in more detail in the next section, these inhomogeneities
are imprinted on the distribution of interstitial black holes at the end of the $p=\rho$ era.  
A fluid of black holes can have two phases, a nonrelativistic gas, and the dense $p=\rho$ state.  The control
parameters which determine which phase the gas is in, are the ratio between the black hole separations and their
average Schwarzchild radius, and the fluctuations in the black hole energy density (which only occur in the $p=0$
phase). The pressure is the order parameter which distinguishes the two phases. At the end of the
$p=\rho$ era we are close to the transition point between these phases. The black hole gas is just dilute enough
to begin expanding like a non-relativistic gas.  However, if there are large density fluctuations, black hole
collisions will lead to the formation of much larger black holes, sending us back to the $p=\rho$ phase.
Thus, we must choose initial conditions with small fluctuations in order to escape the $p=\rho$ phase.
It would be nice to turn this intuitive argument into a numerical estimate of the amplitude of fluctuations, but
we have not yet been able to do so.

In our previous paper we assumed that the universe evolved as a pressureless gas of black holes until Hawking
decay set off a Hot Big Bang era of radiation domination.  The reheat temperature was of order $M^{- {3\over
2}}$, and nucleosynthesis put an upper bound on the fundamental parameter $M$.  However, once we are in a normal
space-time we must consider all the low energy degrees of freedom that the system possesses.  In the context
of string theory, this includes various moduli.   

There are interesting regions of moduli space in which the radius of the compactification is large
compared to the higher dimensional Planck scale, and ${\cal N} = 2$ SUSY breaking is localized on
branes\footnote{This was first emphasized in the work of Horava and Witten\cite{hw}.  The utility of these
regions in cosmology was pointed out in
\cite{tbmcosmo}.}.  In these
regions, the Lagrangian for bulk moduli takes the form 

\eqn{modlag}{{\cal L} = {1\over 2} G_{ij} (\phi) \nabla\phi^i \nabla\phi^j - {\mu^4} V(\phi )}
where $\mu^4 = {M_d^6 }$ and $M_d$ is the higher dimensional Planck mass.  As usual, we work in four
dimensional Planck units.  One might want to consider
$M_d \sim 10^{16} GeV = 10^{-3}$ in order to incorporate Witten's explanation of the ratio between the unification
scale and the Planck scale \cite{hw}.  We will allow $\mu$ to be a free parameter.  In this context, the four
dimensional Planck scale might change as the fields move around on moduli space.  We will assume that this does
not occur, {\it i.e.} that the trajectory on moduli space is orthogonal to changes of the overall radius of
compactification.

It is unnatural to assume that the moduli fields are in their ground state when the energy density is larger than
their potential.  In particular, we may imagine that there is some quantum probability to find a region of
space of size $\mu^{-2}$ over which the fields are homogeneous, and sit at some point on the potential where slow roll inflation can occur.  This is reasonable as long as the particle
horizon is larger than or equal to this size.   If the pressureless black hole gas has evolved for a time $t$
since the end of the $p=\rho$ era, the physical size of the particle horizon is $M a^{3/2}$, where $a = t^{2/3}$.
Thus, the condition for homogeneous scalars is $ Ma^{3/2} \geq \mu^{-2}$.  The energy density of the black holes
at this time is $(M^2 a^3)^{-1}$.  At the time the horizon size reaches $\mu^{-2}$, the energy density in the
moduli is of the same order as that in the black holes.  For an appropriate form of the potential, and initial
values of the scalars, we will enter a period of slow roll inflation.   The black holes will be diluted, and
the energy density will be dominated by the slowly rolling scalars.
For Lagrangians of the form \ref{modlag}, the restoring and frictional forces are of the same order of magnitude
and we naturally get such a period of inflation.  We will call the number of e-folds $N_e$.  Very large values
of $N_e$ are unnatural in this context, but $N_e \sim 10 - 100$ is quite reasonable.  

Reheating of the universe now occurs through decay of the inflaton.  Its lifetime is $o(\mu^{-6})$ and the reheat
temperature $T_{RH} \sim \mu^3$.

The parameters characterizing our cosmology are $M, a, \mu$ and $N_e$.  They are related by $M^2\mu^4 =
a^{-{3}}$.  There are several important constraints on the range of these parameters.  The first is a
matter of principle:  it does not make sense to talk about low energy degrees of freedom like scalar fields,
until the normal cosmology is well established, in other words, when $a$ is large enough.  The $p=\rho$ phase has
no such excitations.  The boundary value of $a$ at which conventional low energy analysis becomes valid is surely
no smaller than $a=10$.  Thus \eqn{ineqa}{ M^2 \mu^4 \leq 10^{-3}}.

The other bounds come from phenomenological requirements.  We will discuss them after we have explained the
nature of the fluctuation spectrum in our model.

\section{Scale Invariant Fluctuations}

In our previous paper we calculated the fluctuations generated during the $p=\rho$ phase by using quantum field
theory expanded around the $p=\rho$ background.  We pointed out there, that this was incorrect.  The $p=\rho$
fluid does not support local field theoretic excitations.  Here we will rederive the scaling law for the
fluctuations using only a symmetry hypothesis.   Like all FRW cosmologies with a single component fluid, the
$p=\rho$ universe
\eqn{prhometric}{ds^2 = - dt^2 + t^{2/3} d{\bf x}^2}
possesses a conformal Killing symmetry
\eqn{killinga}{t \rightarrow \lambda t}
\eqn{killingb}{{\bf x} \rightarrow \lambda^{2/3} {\bf x}}
For {\it e.g.} radiation, it is clear that the analogous conformal symmetry is not a symmetry of the physics. As
we go back in time, the radiation gas gets hotter and irrelevant operators will become important and destroy the
symmetry.  We believe that the dynamics of the dense black hole fluid {\it is} conformally invariant.  The
process of merging black holes has an obvious self-similarity at any scale.  We cannot identify any observables
of this system which detect the change in scale.   Of course, we cannot prove this symmetry conjecture until we
have a microscopic quantum description of the $p=\rho$ fluid.   However, since the conjecture leads to
interesting results, we find it reasonable to leave its justification to future work.

The origin of the inhomogeneous fluctuations in the black hole fluid, are the inhomogeneities which we insert
into the initial conditions.  We have argued that these must be small, in order for the normal region to survive.
The observable fluctuations in the normal region are encoded as fluctuations , $\delta \rho_{BH}$, of the sizes
and positions of the interstitial black holes.   Given the necessary assumption that the initial fluctuations are
small, these are related to the initial shape of the fractal and the matter distribution in it by a linear
formula:

\eqn{flucta}{\delta \rho_{BH} ({\bf x},T) = \int_1^{T} ds \int {d^3 y} f(T,s, {\bf x - y}) \rho_F (s, {\bf y})}

$\rho_F$ is the energy density fluctuation in the fractal, and $f$ is a transfer function.   We will assume that
this transfer function falls off at large separation, in a time independent way.   That is, its Fourier
transform approaches a constant at small ${\bf k}$. Note that
\ref{flucta} takes the form of a convolution, because of the translation invariance of the $p=\rho$ background.
The range of time integration is from something of order the Planck time, up to the end of the $p=\rho$ era, at
time $T$.

We will assume that the fluctuations $\rho_F$ have a Gaussian distribution.  At present we have no argument for
this, because details of the distribution depend on microphysics of the $p=\rho$ fluid.   This translates into a
Gaussian distribution for the primordial fluctuations $\delta\rho_{BH}$.  The low $k$ value for the two
point correlation of the latter variable is given by
\eqn{fluctb}{<\delta\rho_{BH} (k,T) \delta\rho_{BH} (-k, T)> \sim \int_1^T ds ds^{\prime} <\rho_F (k, s) \rho_F
(-k, s^{\prime}) > \equiv \int_1^T ds ds^{\prime} G(k, s,s^{\prime}) }

$\rho_F (x,s)$ is the probability density for fluctuations in the fractal density.  Thus $\int d^3 x \rho_F$
should be invariant under the conformal symmetry of the background.  This leads to the scaling law:

\eqn{scale}{G(\lambda^{-2/3}{\bf k}, \lambda s, \lambda s^{\prime}) = G({\bf k}, s, s^{\prime} )}

When we use this scaling relation for $\lambda \sim k^{3/2}$ and small $k$, the left hand side of the equation
may be evaluated at sub-Planckian times, even though the function itself is evaluated at times larger than the
Planck scale.  We interpret this as an analytic extrapolation of the functional form which satisfies the scaling
relation.  No physical time scale is ever taken to be sub-Planck scale.  By changing variables, and using the
scaling relation, we can write

\eqn{fluctc}{<\delta\rho_{BH} (k,T) \delta\rho_{BH} (-k, T)> \sim {1\over k^3} \int_{k^{3/2}}^{k^{3/2}T } ds
ds^{\prime} G(1, s, s^{\prime})}

The lower limit of integration is effectively zero.  For $k^{3/2} T \geq 1$, the integral will be independent of
the upper limit of integration, if the function $G$ falls off sufficiently rapidly at large times.  In our
previous paper $G$ was a Green's function of the massless Klein-Gordan equation at finite momentum and the
integral was finite.  Here we need only assume that the integral converges
at infinity.   We then predict a ${1\over k^3}$ spectrum for the two point function, which means that the power
spectrum is constant.  {\it Thus, over a certain range of scales, holographic cosmology predicts an exactly scale
invariant spectrum.}  This is to be contrasted with slow roll inflation, which generally predicts small
deviations from scale invariance: the power spectrum is neither exactly constant, nor is it exactly a power law.

In this derivation it appears that the range of comoving scales over which scale invariance holds, ranges from
the Planck scale to the coordinate size of the horizon at the end of the $p=\rho$ era.  However, we cannot really
trust these arguments down to the Planck scale, since the $p=\rho$ fluid is only an approximation to the correct
quantum cosmology when the number of states is large.  We will assume that for $k < .1$, we can use the $p=\rho$
scaling law.   Note that this system might have correlations on scales much larger than the particle horizon.
These could come from the structure of the original fractal\footnote{In his lecture at the Nobel Symposium, one
of the authors, TB invoked such large scale correlations in order to avoid using inflation.  The argument of
this paragraph shows that any such attempt is doomed to failure.}.  However, our derivation shows that scale
invariance persists only up to the horizon scale.   Past this point, the upper endpoint of integration is small
and if the correlation function is smooth, the $k^{-3}$ behavior is replaced by a constant.  Thus, the power
spectrum dies rapidly at small $k$, once $k$ is smaller than the inverse horizon size.  It is possible that this
behavior has observable consequences.

\section{Parameter Fitting and Phenomenology}

 Nucleosynthesis requires $\mu^3 \geq 10^{-22}$.  
We have shown that the fluctuation spectrum generated in the $p=\rho$ era is scale invariant
over a range of scales whose coordinate size is between $10 $ and $M^{2/3}$ at the end of the $p=\rho$ era.
We get two inequalities by requiring that the physical size of this range today encompasses the range of scales
over which approximately scale invariant fluctuations have been observed.  The largest scale at which we have
scale invariant fluctuations is the horizon scale at the end of the $p=\rho$ era.  The physical size of this
scale today is 

\eqn{rcorr}{R_{corr} = 10^{-29} {Ma \over \mu} e^{N_e} R_{Now} = 10^{-29} M^{1/3} \mu^{-{7\over 3}} e^{N_e}
R_{Now}}, where we have included the stretching due to the non-relativistic black hole era, the inflationary era,
and the period from reheating to today.  To fit the data, we need
\eqn{fita}{R_{corr} \geq R_{Now}}
and

\eqn{fitb}{10 M^{- {2\over 3}} R_{corr} \leq 10^{-3} R_{Now}}.  Here we have insisted on scale invariance from
scales on the size of our current horizon, down to the scales which are just going nonlinear today.  In principle
we could also insist on scale invariance down to galactic scales, which would replace $10^{-3}$ by $10^{-5}$.

This can be summarized by two inequalities for the number of e-foldings:

\eqn{ineqb}{10^{29} M^{-{1\over 3}} \mu^{7/3} \leq e^{N_e} \leq 10^{25} M^{1/3}\mu^{7/3},}
which also implies a bound on $M$
\eqn{ineqc}{ M^{2/3} \geq 10^4}

Actually there is a loose first principles argument which also suggests that $M$ should be large.  The $p=\rho$
description of the early universe only becomes valid when the particle horizon is quite a bit larger than the
Planck scale so that there are enough degrees of freedom to justify a statistical description.   Furthermore, $M
\sim {1\over
\epsilon^2}$, where $\epsilon$ is the fraction of the initial volume occupied by the
fractal.  The most probable initial conditions are those which have the smallest values of $\epsilon$ consistent
with survival of the normal regions.  Thus, $M$ is expected to be large in Planck units.

To get some idea of the nature of these bounds, we examine some extreme limits.  When $M$ takes on its minimal
value $10^{6}$, and $a$ is $10$, $\mu \sim 10^{15} GeV$.  $N_e$ is then constrained to be $18 {\rm ln} 10
\sim 41$.  Here $\mu $ is more or less compatible with what one would expect from a GUT scale compactification.
The inflationary energy density fluctuations are of order $10^{-8}$ and cannot account for what is seen in the
CMB.  However, the scale invariant fluctuations left over from the $p=\rho$ era, might be of large enough 
amplitude (we do not know how to calculate the amplitude).  The reheat temperature is of order $10^7$ GeV.
This is compatible with a variety of mechanisms for baryogenesis.  In this range of parameters there are few
constraints on the scale of SUSY breaking, or the nature of dark matter.  The most disturbing thing about
this corner of parameter space is the tight constraint on the number of e-foldings.

The other extreme limit is $\mu \sim 10^{-22/3}$, where the reheat temperature is just at the nucleosynthesis
scale.  Again choosing $a=10$, $M$ is then
$10^{79/6}$.  The number of e-foldings is bounded between $17$ and $31$.  Clearly in this case the inflationary
fluctuations are completely negligible.  We need a very low energy mechanism (like Affleck-Dine\cite{ad}) for
baryogenesis.

\section{Conclusions}

We have now presented a complete cosmology, which solves all of the classic cosmological conundra in a manner
different from inflationary models.  The most important of these is the horizon problem, which actually has
two parts.  The approximate homogeneity of the universe arises from the properties of the $p=\rho$ state.
This state is both homogeneous, and extremely robust.  Large fluctuations away from it\footnote{These large
fluctuations must be imprinted on the initial conditions.  They are the most probable initial conditions which do
not become a $p=\rho$ state.}, which ``attempt" to create a normal region of the universe, must be approximately
homogeneous in order to avoid recollapse into the dense black hole fluid.  The flatness problem is also solved
by appealing to maximal entropy initial conditions.    

The second part of the horizon problem is the correlation between fluctuations on the scale of our current
horizon.   In our model these correlations all arise during the $p=\rho$ phase, and are stretched to the current
horizon size by a short burst of inflation.  There is no region of the parameter space of our model in which
conventional de Sitter fluctuations make a sizable contribution to the observable fluctuations in the CMB.
Our spectrum is exactly scale invariant ($n_S = 1$) and extends over a limited range of scales.  In principle,
for some values of the parameters, the cutoff of the spectrum could have observable consequences.  It might be
related to the apparent lack of power in the CMB fluctuation spectrum at low $L$ values.

There is no monopole problem in our model.   Primordial monopoles are large black holes and these are inflated
away .   Even at the extreme reaches of our parameter space, the reheat temperature is $\leq 10^7$ GeV .  The
universe never undergoes a GUT phase transition and there is no production of monopoles by the Kibble mechanism. 

The observable radiation entropy of our universe is created in the decay of the inflaton, in a more or less
conventional manner.  The reheat temperature of the universe is low throughout our parameter space, which rules
out models of baryo or lepto genesis at the unification scale.  Over much of our parameter range, weak scale
baryogenesis is also ruled out.   Affleck-Dine baryogenesis or baryogenesis in inflaton decay (this is more
difficult and requires relatively large baryon number violating couplings in the low energy theory) are the
mechanisms preferred by our model.   

Our cosmology may also put constraints on the theory of Dark Matter.  For a portion of the allowed range of
parameters, the reheat temperature is too low for supersymmetric dark matter.  Superpartners are never in thermal
equilibrium so the conventional calculation of their relic abundance is incorrect.  For some of the range
of parameters, they are too heavy to be produced in inflaton decay at all.  Even when they can be produced, their
relic abundance will depend on many details of the decay.   An alternative candidate for dark matter is an axion.
Recent calculations\cite{fpt} indicate that an axion with decay constant $10^{14} ({6 {\rm MeV} \over T_{RH}})$
GeV is a good dark matter candidate in these models.

There are two interesting ways to compare our theory to standard inflation.  The first is an assessment of the
relative probability of the initial conditions which lead to the two models.   Consider an equal area time slice
on which the size of the particle horizon is $L \gg 1$.  The entropy of the initial conditions which lead to 
a fractal which breaks away from the $p=\rho$ state, is of order $L^2$.   Now consider initial conditions for
inflation on the same slice.  The system must be described by field theoretical degrees of freedom, whose entropy
is bounded by $L^{3/2}$.   Thus, the initial conditions for holographic cosmology are more probable.

On the other hand, our bounds on $M$ and $N_e$ coming from the requirement that $p=\rho$ fluctuations are
responsible for the CMB, are not sacred.  It might turn out that microphysics fixes $M \sim 10^3$ .  In this case
$\mu$ is in the right range to generate observable inflationary fluctuations.   Depending on the number of
e-folds, and on the amplitude of $p=\rho$ fluctuations, they may also contribute to the CMB.  Such a scenario
requires a larger number of e-folds, and so might be harder to achieve .  Furthermore, such low values of $M$
do not seem probable.   The question can only be resolved by understanding the microscopic physics which
underlies the $p=\rho$ state.  This will lead to a calculation of $M$.

\section{Appendix}

The fundamental insight which  ties geometry to quantum mechanics is the holographic entropy bound for causal
diamonds.  This bounds the entropy associated with a causal diamond in any Lorentzian geometry.  If the causal
diamond is small enough ({\it e.g.} small compared to the AdS radius in an asymptotically AdS space-time)
this bound is finite.  Although the entropy of thermal density matrices of infinite systems is often finite, the
concept of thermal density matrix requires a preferred Hamiltonian.  In a local region of space-time in a theory
of quantum gravity, there are no such privileged operators\footnote{This is the Problem of Time, to which we
will return momentarily.}.  Thus, we believe that the only reasonable interpretation of the finite entropy bound
for a causal diamond is that the quantum theory describing this submanifold of space-time has a finite number of
degrees of freedom.

This hypothesis introduces a notion of locality into the quantum theory of gravity (QG), which is compatible
with the holographic principle.  The intrinsic non-locality of QG only manifests itself when we consider what
happens to larger and larger causal diamonds.  In asymptotically flat space-time the holographic bound converges
to infinity as the time-like separation betwen the tips of the causal diamond is taken to infinity.  In
asymptotically AdS space-time, the bound reaches infinity when the time-like separation is finite.  In
asymptotically dS space-time, the bound saturates at a finite value even when the separation is infinite.  This
motivates the conjecture\cite{tbf} that the number of states in dS space is finite.  In Big Bang cosmologies,
causal diamonds stop at a finite point in the past, and are replaced by causal cones.  This sheds light on the
nature of the Big Bang singularity.  If the number of degrees of freedom describing physics inside the particle
horizon is finite, and gets very small as the Big Bang is approached, it is easy to understand why effective field
theory, or weakly coupled string theory approximations break down, despite the fact that the quantum system is
completely under control.  Thus we view the Big Bang as the point at which the Hilbert space describing local
physics has a certain minimal size.

Implicit in this description are two principles which are somewhat at variance with conventional notions of how
physics works.  The first is an explicit breaking of TCP invariance.  We claim that the definition of a quantum
Big Bang cosmology has a built in arrow of time, corresponding to the fact that the area of the particle horizon
increases.   This also suggests a special time slicing: we choose slices such that the area associated with the
causal past of each point on a given slice, is the same.  On these slices
generic initial states obviously saturate the entropy bound.  Since a homogeneous $p=\rho$ fluid can saturate the 
entropy bound, we argue that it is the correct coarse grained description of the quantum evolution of generic
initial states.   As we described at length in \cite{holocosmo} this leads to a solution of the horizon and
flatness problems which is different from that given by inflation.

Here we want to give a brief summary of our attempts to find a microscopic quantum description of the $p=\rho$
fluid. The key question is, ``What is the Hamiltonian?".  We claim that this is the classic relativist's
question called The Problem of Time.   In a generally covariant theory which has no asymptotic boundaries (or if
we want to describe physics in a quasi-local fashion), the definition of time, and thus of the Hamiltonian, must
be observer dependent - that is, non-gauge invariant.   In addition, there is no reason to expect any given
observer to have a time independent Hamiltonian.  Finally, we note that we may expect that in general the time
evolution operators corresponding to different observers will not commute with each other.  This is the basis of
the notion of {\it observer complementarity}, which may potentially resolve the puzzle of black hole evaporation.

Our description of quantum cosmology, breaks the Hilbert space of a given observer down into a nested tensor
factorization.  Smaller factors correspond to earlier times.  An ordered sequence of unitary operators, $U_n =
e^{i H_n}$ describes the dynamics.   Each small factor space has its own subsequence of unitaries, and all of
these descriptions must be compatible with each other.  For a single observer this constraint is easy to satisfy.
$H_k$ with $k < N$ in the Hilbert space corresponding to some late time $N$ steps from the Big Bang, should not 
couple the degrees of freedom that act on the tensor factor ${\cal H}_k$ to the rest of the system, and it should
agree with the time evolution described in this smaller space  (Fig. 3).

The constraint that different Hilbert spaces, corresponding to observers following neighboring trajectories, have
consistent dynamics, is harder to describe, and has so far resisted our attempts to solve it.  For this reason,
the proposal we are about to make for the microscopic $p=\rho$ dynamics cannot yet be shown to be a consistent
quantum cosmology.
\vfill\eject

\includegraphics[width=5in]{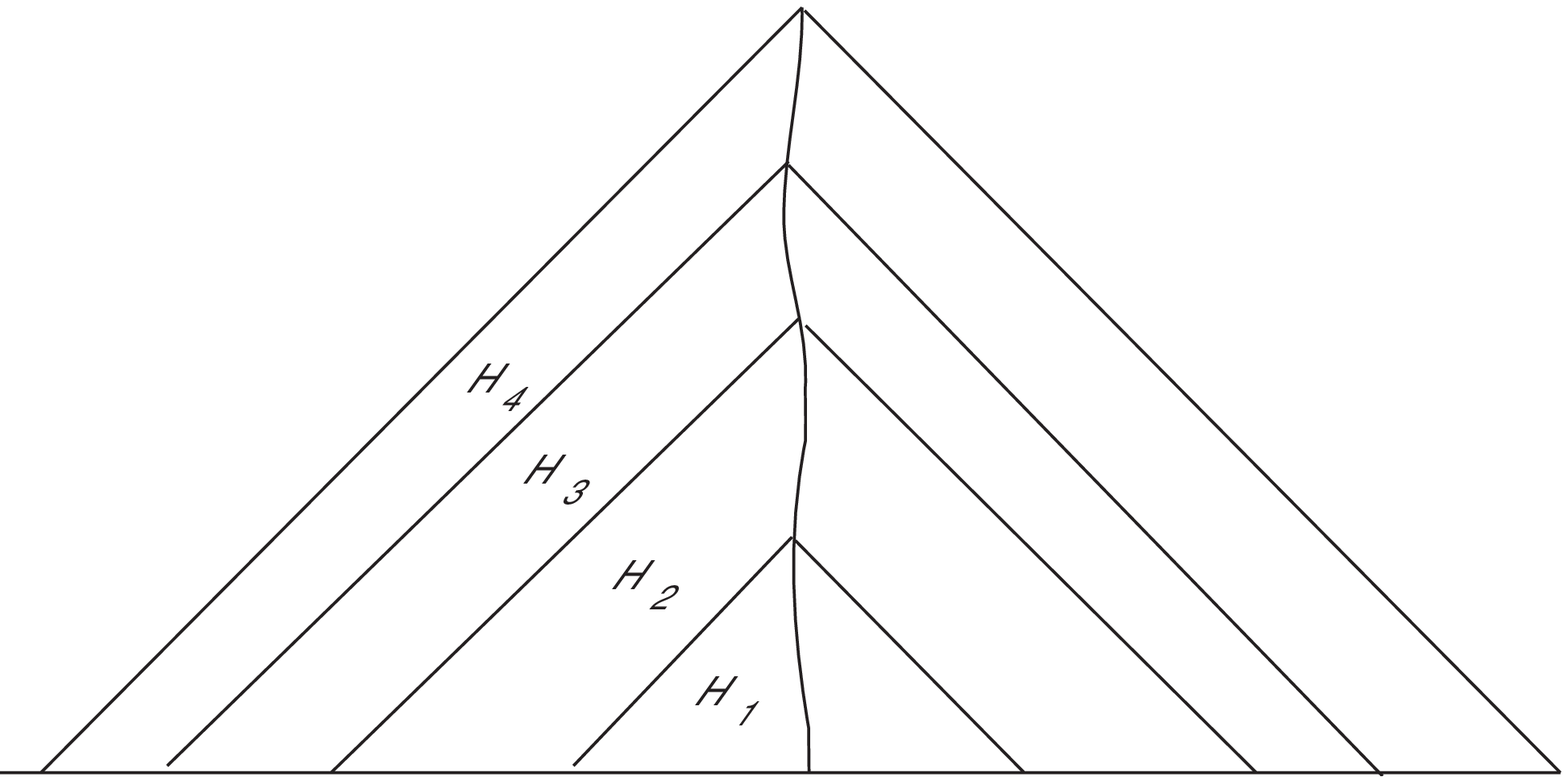}

\centerline{Figure 3: Nested Hilbert Spaces Define Particle Horizons}
\centerline{\ \ \ \ Along a Timelike Trajectory}

\vskip.4in
Our proposed dynamics is essentially, that at each time, the Hamiltonian $H_n$ be a random operator on the
Hilbert space on which it acts.  To describe this more precisely, consider the spinor variables $S_a (n)$ which
were suggested in \cite{susyholo} as the appropriate quantum mechanical description of holographic screens.
These satisfy
$\{S_a (k), S_b (l) \} = \delta_{ab}\delta_{kl}$
for $k,l$ between $1$ and $N$.  They generate the operator algebra on ${\cal H}_N$.

We begin with a general quadratic Hamiltonian $S_a (l) S_b (k) H^{ab}_{kl}$  The matrix $H$ is antisymmetric
under exchange of $(a,k)$ with $(b,l)$.   If we choose this from the Gaussian random ensemble of such matrices,
then for large $N$ there is a large range of eigenvalues near zero, for which the spectrum is linear.  Thus, for a
random matrix the low energy spectrum is that of a $1+1$ dimensional conformal field theory and satisfies
the energy entropy density relation $\sigma \sim \sqrt{\rho}$.  We would like to identify these CFT quantities as
the bulk spacetime energy and entropy densities of an FRW universe.  This of course works only if the equation of
state in space-time is $p=\rho$.

This property of random quadratic Hamiltonians persists beyond the quadratic approximation.  Most perturbations
of the free $1+1$ fermion fixed point are irrelevant.   There are marginal four fermi operators, but for one sign
of the coupling, they are marginally irrelevant.  Thus, over roughly half the volume of the parameter space
of random Hamiltonians, we predict a universal spectral behavior with the required relation between energy and
entropy. 

Note that our prescription is to pick a {\it different} random Hamiltonian at each point in time.  Thus, although
the energy spectrum quickly settles down to a universal form, there is no sense in which the system settles down
to the ground state of any particular Hamiltonian.  Different members of the random ensemble will not commute
with each other, and the system will explore its full Hilbert space.   We claim that this is precisely the sort
of behavior one might expect for a "dense black hole fluid":  smooth coarse grained energetics, but a completely
random and therefore maximally entropic state vector.  

In order to prove that this microscopic description is indeed the $p=\rho$ fluid discussed in the text we must do
two things.  First we must show that the prescription can be extended in a homogeneous and isotropic way, which
satisfies the consistency conditions of quantum cosmology.   That is, we must introduce other sequences of nested
Hilbert spaces corresponding to observers on nearby trajectories.  We give them precisely the same random
dynamics as the original sequence.  Two nearest neighbor sequences of Hilbert spaces share almost all of the same
degrees of freedom, corresponding to the fact that backward lightcones from points on the two trajectories have
a large intersection.  A map between the two operator algebras must be found, which is consistent with the
dynamics.  The hypotheses of homogeneity and isotropy imply that the same map is used for any pair of nearest
neighbor trajectories.   We have not yet found this map.

Another check of our model would be a derivation of the scaling law relating the energy density to the total
entropy in the $p=\rho$ FRW cosmology, from the scaling laws of large random matrices.

Work on these problems is in progress, but we have not yet accomplished these goals.

\acknowledgments We would like to thank the KITP for it's hospitality during the "String Theory and Cosmology"
workshop when this paper was being written and the participants for useful discussions, in particular Rami
Brustein , Sean Carroll, Nemanja Kaloper, and Gabriele Veneziano. TB would like to thank Joel Primack for discussions on the
properties of the spectrum of fluctuations.  He would also like to thank the organizers of the Nobel Symposium,
and in particular Ariel Goobar, for inviting him to present this work in that venue, and for help with his
presentation.

The research of W. Fischler is based upon
work supported by the National Science Foundation under Grant No.
0071512. The research of T. Banks was supported in part by DOE
grant number DE-FG03-92ER40689.



\newpage
\begin{thebibliography}{19}








\bibitem{holocosmo} T.~Banks, W.~Fischler,
{\it An Holographic Cosmology}, hep-th/0111142

\bibitem{fsb} W.~Fischler, L.~Susskind,
 {\it Holography and Cosmology,} hep-th/9806039; 

R. Bousso, {\it Holography in
general space-times,} JHEP 9906, 028 (1999) hep-th9906022

E.~Flanagan, D.~Marolf, R.~Wald, {it Proof of classical versions of the Bousso entropy bound
and of the generalized second law,} hep-th/9908070


\bibitem{g} G.~Veneziano, Phys. Lett. {\bf B454}, 22 (1999)

R.~Easther, D.~Lowe, Phys. Rev. Lett. {\bf 82}, 4967(1999)

D.~Bak, S.~Rey, {\it Cosmic holography,} hep-th/9902173

N.~Kaloper, A.~Linde, Phys. Rev. {\bf D60} 103509(1999)

R.~Brustein, G.~Veneziano, Phys. Rev. Lett.{\bf 84} 5695(2000)


\bibitem{bfmcosmo} T.~Banks, W.~Fischler,
{\it M-theory observables for cosmological spacetimes,}
hep-th/0102077.

\bibitem{ad} I.~Affleck, M.~Dine, 
{\it A new mechanism for baryogenesis,} Nucl.\ Phys.\ {\bf B249}, 361 (1985)

\bibitem{fs} W.~Fischler, L.~Susskind,
 {\it Holography and Cosmology,} hep-th/9806039

\bibitem{israel} W.~Israel, Nuovo Cimento {\bf 44B}, 1 (1966)

\bibitem{hw} P.~Horava, E.~Witten,
{\it Heterotic and type I string dynamics from eleven dimensions,} Nucl.\ Phys.\ {\bf B460}, 506 (1996)
E.~Witten, {\it Strong Coupling Expansion of Calabi Yau Compactification}, Nucl.\ Phys.\ {\bf 471}, 135 (1996).
\bibitem{tbmcosmo} T.~Banks, {\it Remarks on M-theoretic Cosmology}, hep-th/9906126.

\bibitem{fpt} P.~Fox, A.~Pierce, S.~Thomas, {\it Probing a QCD String Axion with 
Precision Cosmological Measurements}, talk
presented at KITP workshop on 
Superstring Cosmology, Oct. 2003,     
http://online.kitp.ucsb.edu/online/strings03/thomas/ 

\bibitem{tbf} T.~Banks, {\it Cosmological breaking of supersymmetry or little lambda goes back to the future II}, hep-th/0007146

W.~Fischler, {\it Taking de Sitter seriously}, Talk given at {\it Role of scaling laws in physics and biology (Celebrating the 60th birthday of Geoffrey West)}, Santa Fe, Dec. 2000


\bibitem{susyholo} T.~Banks,
{\it SUSY and the holographic screens,} hep-th/0305163



\end{thebibliography}
\end{document}